\documentclass[letterpaper]{article} 
\usepackage{aaai25}  
\usepackage{times}  
\usepackage{helvet}  
\usepackage{courier}  
\usepackage[hyphens]{url}  
\usepackage{graphicx} 
\usepackage{natbib}  
\usepackage{caption} 
\usepackage{algorithm}
\usepackage{algorithmic}
\usepackage[table]{xcolor}
\usepackage{pifont}
\usepackage{array}

\usepackage{longtable}
\usepackage{booktabs}
\usepackage{makecell}
\usepackage{tabularx}
\newcolumntype{L}[1]{>{\raggedright\arraybackslash}p{#1}}

\usepackage{newfloat}
\usepackage{listings}
\DeclareCaptionStyle{ruled}{labelfont=normalfont,labelsep=colon,strut=off} 
\floatstyle{ruled}
\newfloat{listing}{tb}{lst}{}
\floatname{listing}{Listing}
\definecolor{LightGrey}{gray}{0.85}
\definecolor{LightOrange}{RGB}{255, 230, 200}
\definecolor{LightBlue}{RGB}{200, 220, 255}
\newcommand{\NotSpecified}{\cellcolor{LightGrey}Not Specified}
\newcommand{\Yes}{\cellcolor{LightOrange}Yes}
\newcommand{\No}{\cellcolor{LightBlue}No}
\title{User Privacy and Large Language Models: \\ An Analysis of Frontier Developers’ Privacy Policies}
\author {
    Jennifer King\textsuperscript{*}, Kevin Klyman\textsuperscript{*}, Emily Capstick, Tiffany Saade, Victoria Hsieh
}
\affiliations{
    Stanford University\\
    Stanford, CA, USA\\
    Correspondence to kingjen@stanford.edu and kklyman@stanford.edu
}

\usepackage{bibentry}

\begin{document}

\maketitle

\begin{abstract}
Hundreds of millions of people now regularly interact with large language models via chatbots. Model developers are eager to acquire new sources of high-quality training data as they race to improve model capabilities and win market share. This paper analyzes the privacy policies of six U.S. frontier AI developers to understand how they use their users' chats to train models. Drawing primarily on the California Consumer Privacy Act, 
we develop a novel qualitative coding schema that we apply to each developer’s relevant privacy policies to compare data collection and use practices across the six companies. 
We find that all six developers appear to employ their users' chat data to train and improve their models by default, and that some retain this data indefinitely. Developers may collect and train on personal information disclosed in chats, including sensitive information such as biometric and health data, as well as files uploaded by users. Four of the six companies we examined appear to include children's chat data for model training, as well as customer data from other products. 
On the whole, developers' privacy policies often lack essential information about their practices, highlighting the need for greater transparency and accountability.
We address the implications of users' lack of consent for the use of their chat data for model training, data security risks arising from indefinite chat data retention, and issues with training on children's chat data. We conclude by providing recommendations to policymakers and developers to address the data privacy challenges posed by LLM-powered chatbots.


\end{abstract}

%

\section{Introduction}

For over three decades, privacy policies have provided internet consumers with details about the personal data companies collect from them and how they use this data. As a method of public communication, they are deeply flawed: dense, difficult to read, and typically written in convoluted language that is challenging to understand without a law degree. Consumers rarely read them, let alone correctly understand the practices they document \citep{Turow03072018}---if they tried, it would take hundreds of hours \citep{mcdonald2008cost}. 
\newline
\indent Nevertheless, as artificial intelligence systems powered by large language models (LLMs) proliferate, and consumers interact with LLMs through chat interfaces---chatbots---privacy policies remain the core resource for consumers seeking answers about how the conversations they share with chatbots are collected and processed. While researchers and others have voiced concerns about vast troves of data scraped from the internet to train LLMs \citep{ leffer2023personalinfo, chen2024aidata, solove2025greatscrape}, fewer to date have focused on the collection and use of data directly from consumer interactions with LLMs. This is due in part to the rapid transformation of the industry. Before 2022 there were very few users of LLMs, but the release of ChatGPT—the fastest-ever growing consumer application—catalyzed a meteoric rise in consumer use, with 700 million ChatGPT users as of August 2025 \citep{openai2025users}. 
\newline
\indent Developers have been consistently criticized for privacy issues related to scraping data, including personal data, from across the internet to build LLMs \citep{PAULLADA2021100336, elazar2024whatsbigdata}. This new vector for the collection and use of personal data presents a challenge to global data protection regulation frameworks, as evidenced by the wave of inquiries by data protection authorities worldwide in the wake of ChatGPT's launch \citep{zanfir-fortuna2023}. Personal data is often included in web scrapes without individuals' knowledge or consent, running afoul of existing frameworks like the European Union's General Data Protection Regulation (GDPR). In the U.S., our focus in this paper, the question of privacy protections for personal data collected by or shared with LLM developers is complicated by a patchwork of state level laws and a lack of federal regulation. 
\newline
\indent This paper analyzes the privacy policies of six U.S.-based developers of LLM-powered chatbots to answer basic questions regarding the chat data they collect from their users and how they use this data. More specifically, we investigate whether and how their privacy policies disclose: (i) whether user inputs to chatbots (i.e., prompts and other data) are used to train or improve LLMs; (ii) what sources and categories of personal consumer data are collected, stored, and processed to train or improve LLMs; and (iii) users' options for opting into or out of the use of their chats for training. We ground our analysis of these policies primarily in California's data privacy law, the California Consumer Privacy Act (CCPA), as it is the most comprehensive privacy law in the U.S. and all six of these developers are required to comply with it when serving California consumers.
\newline
\indent We found that all the developers rely upon a web of documents in addition to their primary privacy policies to govern their use of users' chat data. OpenAI, for example, relied on at least six separate policies. Developers’ privacy policies handle LLM training data differently based on source. Some, for example, attempt to strip personal data from scraped data in their training corpus. However, all use user chats (inputs) with their chatbots by default to train their LLMs. Most require that users who wish to avoid having their data used for model training affirmatively opt-out, while others do not offer opt-outs.  Developers that maintain other products on their platforms that collect substantial user data, such as Google and Meta, disclose that they may use their customers' data from other services to train LLMs or personalize chatbot interactions. 

We make three main contributions. First, we provide new insights regarding how chatbots used by millions collect and use their chat data. Second, we identify broader impacts of chat data collection and use: the implications for vulnerable groups such as children, the power of defaults to facilitate data collection, and the risks of indefinite data retention and the use of sensitive personal information to build LLMs. We raise questions regarding whose interests this data collection serves. Third, we advance specific policy recommendations to address these concerns.
 
This paper is organized as follows: the literature review places privacy policies for LLM-powered chatbots in the context of prior work on privacy policies for online platforms and websites. The methodology section lays out the method we used for selecting developers, identifying privacy policies, and building and applying our coding schema. The analysis section compares each policy and describes their provisions on collection and use of chat data. The discussion section discussed the broader implications of our findings. We close with recommendations for developers, policymakers, and researchers.

\section{Background and Relevant Literature}

\textbf{Privacy policies: necessary but not sufficient.} In the United States, Section Five of the Federal Trade Commission (FTC) Act and adoption of the Fair Information Practices both provide the FTC the authority to require that businesses provide notice of their data collection and usage practices to consumers as well as adhere to them \cite{Hoofnagle_2016}. On the internet, this took the form of privacy policies---publicly posted legal notices that describe a business' data practices in detail. This framework, known as notice and consent, places the burden on the consumer to investigate a company’s data practices and make an informed decision about whether to use a service based on them. Researchers have long identified the deficiencies with privacy policies for communicating essential information to consumers. They are typically too long, poorly presented as dense blocks of text, and written in legalese that requires a graduate degree to comprehend \citep{10.5555/3235866.3235868}. Privacy policies lack a standardized format, and do not accommodate variances in consumers’ education levels, native language, and knowledge of or familiarity with the details of how personal data is collected and used across the internet \citep{10.1145/1753326.1753561}. There are limited requirements regarding their readability, and none for their visual or information design that could improve consumers' understanding and awareness of privacy policies and practices. Improvements in a firm's privacy policy design typically occur only after a regulator mandates it via a consent decree or after litigation over a company’s privacy practices \citep{CIPL2019}. 

Despite these drawbacks, in the U.S. a lack of federal omnibus consumer privacy protections means that privacy policies remain a vital tool for tracking what data companies collect from consumers and how they use it. The narrow, patchwork protections extant at the national level largely do not protect the vast majority of data that consumers share with online services, though various state level laws limit companies' use of consumer data, including sensitive personal data such as precise geolocation, health information, and biometric data \citep{iapp2024privacychart}. The U.S. also currently has no federal level AI regulation and only a sparse patchwork of narrow regulations at the state level \citep{IAPP_State_AI}. As a result of this regulatory vacuum, developers of LLM-powered chatbots face few barriers to both scraping data from the internet and using chat inputs from users to build AI systems. 

\textbf{The rise of LLMs and challenges to data privacy.} Given the current dominance of U.S. companies in the market for LLMs, the data used to build these models has global impacts. The FTC has taken action against three companies that used personal information for training AI systems, requiring them to delete both the data and their algorithms \citep{Rite_aid, everalbum, wwatchers}. However, no actions have been taken as of yet against large LLM developers for privacy violations. The copyright claims winding their way through the courts against LLM developers are the clearest pushback to date against data scraping practices \citep{knibbs2024aiCopyright}. Additionally, a lack of transparency requirements for model development make it difficult for members of the public to learn whether and how their personal data is collected and used by AI developers. Some may post information publicly in model training documents, such as model cards, but these are optional, unstandardized, and there are no requirements for disclosing data practices, such as sources of model training data. California's Training Data Transparency Act (AB2013), which takes effect in 2026 and is the leading U.S. effort in this area, will have limited impact on transparency due to its focus on `high level summaries' and a lack of enforceability \citep{shen_disclosure_2025}. As a result, privacy policies remain the primary vehicle for documenting data collection and processing practices that have a direct impact on consumers.

The rise of large language models coincided with heightened interest in data privacy regulation in the U.S. Since the enactment of the CCPA in 2020, 19 additional states passed consumer privacy laws, though California’s remains the strongest in terms of breadth \citep{iapp2024privacychart}. Congress has proposed federal consumer privacy regulations multiple times, most recently in 2024, only to see these bills die in committee. While these state laws do not explicitly regulate AI \citep{king2024rethinking}, most will implicitly impact the practices developers follow when training AI systems using internet and chat data. That these laws do not explicitly regulate data used for AI training may mean that aspects of the LLM training lifecycle, such as the use of chatbot inputs and  outputs in training corpora, are not clearly called out as data sources that require specific protection, such as requiring developers to remove personal data (including sensitive data) from chats prior to using them for model training. 

These potential loopholes are concerning because user interactions with chatbots may be especially prone to including sensitive data \citep{mireshghallah2024trustbotdiscoveringpersonal}. Users can disclose any personal data they wish to a chatbot, and unlike filling out fields in a form or making a search query, chats may disclose far more contextual information about individuals. The conversational give-and-take of a dialogue with a chatbot may not only increase the depth and breadth of personal disclosure, but also include voice recordings, videos, documents, and images uploaded by users. 
Increasingly, chatbot developers are offering users the ability to personalize chat experiences by using “personal information like names, locations, or family members, capturing structured or semi-structured information like user preferences, or modeling interaction patterns, and persistently storing those details to call on later” \citep{sampson2025personalization}.

The principle of purpose limitation is underdeveloped in the context of LLMs. For example, OpenAI posted user chats on the web for indexing by search engines when users shared their chats with others; it removed the feature after public backlash \citep{ars_index_chat}. Our analysis of the Nova chatbot demonstrated that Amazon also posts shared chats for search indexing (albeit with the user's name removed).

\textbf{Dataset accountability.} User chat inputs alone are not enough to create and sustain frontier models, which is why publicly available internet data has been widely scraped to train them. While there is little transparency about the datasets used to train major companies' LLMs \citep{bommasani2023foundationmodeltransparencyindex}, most provide notice that personal data likely exists in their training sets. \citet{osu2025kbpage220} argue that the external collection practices for training data implicate data privacy harms both at the individual as well as the societal level, emanating from the loss of control over personal information when it is scraped, collected, and reused without consent and out of the original context of collection, with no practical way to opt-out. Drawing on the work of Calo \citep{calo}, they argue that “dataset development as a systematic strategy of ingesting personal information” poses two forms of data privacy harms: subjective harms (emanating from unwanted observation) and objective (unanticipated or coerced use of personal information against an individual without first-hand knowledge). Specific privacy concerns with dataset construction include the automation of data collection and the widespread reach of the collection itself, resulting in systematic, mass surveillance of the online sphere across a myriad of contexts that can eventually lead to downstream harms by automated decisionmaking systems.

\textbf{The race for ever more data.} LLMs have become increasingly capable due to an ever-larger amount of training data \citep{kaplan2020scalinglawsneurallanguage}. As a result, firms have raced to scrape as much internet data as possible, leading to training data shortages \citep{villalobos2022will}. Developers admit that they have already scraped as much English-language data as possible from the web, leading them to ink licensing agreements for content providers' proprietary data \citep{schomer2024}. As \citet{longpre2024consent} documented, commons-based data sources are in decline as more websites block web crawlers. The scrape-everything approach means that a huge variety of sensitive data has already been ingested into LLMs, including personal data that may be identifiable, previously obscure and downranked by search engines, or even published in data breaches \citep{10.1145/3531146.3534637}.
\newline
\indent The need for more data—in particular text data—for training and improving LLMs incentivizes developers to collect as much data as possible directly from their own customers. This may lead companies to undermine the data protection principles of data minimization (which calls for data processors to collect only the data necessary for a specific purpose), purpose limitation (which deems that data must be used only for the purpose for which it was collected), and consent (which states that consumers must knowingly assent to the collection and use of their personal data). For example, large platforms that hold vast reserves of data on billions of users have stated that they intend or reserve the right to use data from their non-LLM platforms to train models, raising questions about how such actions will align with data minimization and purpose limitation \citep{burgess2024stopdata}.
\newline
\indent \textbf{Empirical research on privacy and LLMs.} While there is more than a decade of research analyzing privacy policies across a variety of facets \citep{10.1145/3698393}, as well as a growing body of literature analyzing user privacy concerns and expectations with chatbots \citep{Gumusel}, there are few studies to date analyzing the specific privacy practices promulgated by chatbot developers. Surfshark, a security firm, conducted a study quantifying the data collection practices of ten AI chatbots, with Meta AI and Google Gemini topping their list of the most items of data collected of thirty-five items tracked \citep{surfshark2025aichatbots}. \citet{duffourc2024privacy} conducted an in-depth legal analysis of how U.S. and EU legal privacy frameworks govern personal data across the generative AI development lifecycle. They note that the scope of GDPR governance includes the publicly available personal data most commonly scraped from the web, and that processing such data must fall under one of the valid legal bases under GDPR, typically contract, consent, or legitimate interest. In contrast, the scope of the CCPA excludes most publicly available personal information, such as public records or social media profiles. While the authors argue that other U.S. laws, both federal and state, also are unlikely to provide privacy protections to publicly available data, they do not address how these various laws will treat user inputs to chatbots (both conversations and other user-submitted data), chatbot outputs (including responses and reasoning tokens), and chats stored as memory by users.  

\section{Methodology}
\textbf{Company selection protocol.} We analyzed the privacy policies of six of the largest LLM developers in the U.S.: Amazon, Anthropic, Google, OpenAI, Meta, and Microsoft. We chose these companies due to their inclusion as part of the Frontier Model Forum, an industry group created by these companies to share best practices about AI safety and security, which indicates that these companies are at the industry frontier (whether in terms of the capabilities of their models, or in terms of access to compute). By some estimates these companies represent nearly 90\% of market share for LLM-powered chatbots in the U.S., with OpenAI’s ChatGPT accounting for over half the market \citep{statcounter_ai_chatbot_market_share}. We limited our analysis to U.S. companies in order to draw more reasonable comparisons between them.

We did not consider small startups that have more nascent privacy practices (e.g., Perplexity, Together AI, Writer), developers of text-to-image models (e.g., Adobe, Midjourney), firms that have been acqui-hired (e.g., Adept, Character.ai, Inflection), or firms that release open-weight models but (unlike Meta) do not distribute their LLMs through consumer chatbots via their own platform (e.g., Nvidia, Databricks). This narrower scope helps us compare apples to apples and focus on the most prominent chatbot developers. 

We elected to focus on U.S. developers of LLM-powered chatbots because of their dominance of the global AI sector and the fact that the U.S. lacks omnibus consumer privacy regulation similar to the GDPR. The CCPA, the U.S.’ most comprehensive data privacy law, does not prohibit the collection of personal data. Companies subject to the CCPA must provide California residents six data privacy rights: right to limit the use and disclosure of their sensitive personal information; right to opt-out of sale or sharing of personal information; right to correct inaccurate information; right to know what information has been collected about them; right to delete personal information, and a right to equal treatment when exercising these rights \citep{cppa2025faq}. Companies must post their privacy policies online, linked from their homepage (or within their mobile apps), and disclose within their policy: a list of the categories of personal information collected about consumers over the previous twelve months; the sources for each category; the business purpose for collection, selling, or sharing; and the third parties to whom they disclose personal information, if any \citep{calccpa1798130}.
To the extent that AI developers are engaging in practices that maximize the collection of personal data from users, we hypothesized that analyzing U.S.-based practices would demonstrate them with greater clarity. 

\textbf{Policy selection protocol.} We analyzed the privacy policy that applies to the consumer-facing chatbot interface for Amazon (Nova), Anthropic (Claude), Google (Gemini), Meta (Meta AI), Microsoft (Copilot), and OpenAI (ChatGPT). We chose to analyze these systems because: (i) chatbots are how consumers most frequently interact with LLMs; (ii) consumer-facing chatbots are used to generate a large amount of content that has a significant societal impact; and, (iii) each company has its own chatbot service (whereas not each company operates an API or a cloud platform). Nevertheless, this decision has drawbacks as these chatbots are black boxes and many users make use of them not through chat interfaces directly but while integrated into platforms such as Instagram (for Meta AI), Slack (for Claude), or Amazon Web Services (for Nova).

The CCPA describes the purpose of privacy policies as providing consumers ``with a comprehensive description of a business's online and offline information practices" \citep{cppa_sec7011}. Our evaluation of the clarity and consistency of these policies is based upon this articulation. For each company, we identified the relevant privacy policies by (i) beginning with the company’s main privacy policy for all its services and (ii) following links from that policy to find other sub-policies that relate to and cover its LLM-powered chatbot. We found that some of these policies were not adequately specific to the service we were examining, so we also created a new account with each chatbot and collected the policies that were linked in user onboarding, the chatbot interface, and user settings. This resulted in a collection of policies for each company: a ``parent’’ privacy policy (i.e., the company’s most recent overarching privacy policy as of May 2025) and ``branch’’ sub-policies, discussed in the subsequent section.\footnote{Prior to publication, Anthropic announced a policy update from opt-out for model training to opt-in, which we account for.} In some cases, the sub-policies were not policy documents but FAQs.
\newline
\indent \textbf{Coding method for policies.} We conducted an inductive coding analysis of each company’s primary privacy policy as linked from the company’s homepage as of May 2025, as well as relevant sub-policies linked within it or from the chat interface. We developed a coding schema based on the data protection principles articulated in the CCPA, as well as selected facets from GDPR. Additionally, we created categories to code for LLM-specific usage practices that existing data protection frameworks do not capture, such as the collection and use of training data. For this stage of research, it made more sense to manually implement an inductive coding strategy rather than rely on automated methods for parsing privacy policies, particularly since we were interested in identifying emergent data collection and use practices specific to AI development that may not be documented in previous research projects. We coded for specific provisions of the privacy policy including types of first party data collection and use, methods for data retention, storage, or transfer, the legal basis claimed for data processing, any third-party data sharing or sales, and which data is excluded. 
\newline
\indent Coding was split between the team members, and each member’s coding was reviewed by a different team member for consistency in applying the schema. We conducted our coding and analysis using AtlasTI. We intentionally did not use LLMs to conduct our analysis due to concerns regarding lack of reliability and inaccuracies from hallucinations. 
\newline
\indent \textbf{Limitations}. Our research is limited by our focus on our small sample of major developers. There are some differences in business models across developers  (we include multiple cloud providers, some startups, and some companies that release open models) that could impact the scope of the data these companies collect. We also do not include and code every possible sub-policy offered by each developer. These policies are also moving targets and change often. Other than Anthropic, our analysis focuses on policies as they stand at a specific moment in time (May 2025) and does not attempt to track the evolution of chatbot-relevant privacy policies since their introduction in 2022. Nor did we seek to include all possible documents that may describe a company's data-related practices, including model cards, as we discuss below. Our primary assumption in selecting the documents we coded was that they either should be linked directly from the privacy policy, or available directly from the chat interface, and that users should not have to work to identify all of the possible other sources of this information on a developer's platform. In future research, we intend to analyze policies longitudinally to track their evolution and broaden the scope of the firms we target. 

\begin{table*}[t]
\centering
\small
\setlength{\tabcolsep}{8pt} 

\resizebox{\textwidth}{!}{
\begin{tabular}{|l|c|c|c|c|c|c|}
\hline
 \textbf{Data Privacy Practice} & \textbf{Amazon} & \textbf{Anthropic} & \textbf{Google} & \textbf{Meta} & \textbf{Microsoft} & \textbf{OpenAI} \\
\hline
Chat input used for training by default & \NotSpecified & \Yes& \Yes & \Yes & \Yes & \Yes \\
\hline
Mechanism to opt out of chat training & \NotSpecified & \Yes& \Yes & \NotSpecified & \Yes & \Yes \\
\hline
Chat data retained indefinitely & \Yes & \No& \No & \Yes & \No & \Yes \\
\hline
Chatbot personalization features & \NotSpecified &  \NotSpecified & \Yes & \Yes & \Yes & \Yes \\
\hline
Allows accounts for children 13-18 & \No & \No & \Yes & \Yes & \Yes & \Yes \\
\hline
\end{tabular}
}  

\caption{Comparison of Company Data Privacy Practices. Each column is a chatbot developer and each row is a practice as captured by their privacy policies. `Yes' signifies that the company does implement that practice based on the evidence we surface, `No' indicates it does not, and `Not Specified' indicates its policies do not specifically address the practice or that the wording was ambiguous. \textit{Chat input used for training by default} refers to whether firms use inputs to chatbots for training their AI systems by default. \textit{Mechanism to opt out of chat training} refers to whether firms offer users a clear opt-out mechanism for inclusion of their chats in training corpora. \textit{Chat data retained indefinitely} refers to whether firms retain chat inputs/outputs indefinitely or delete them periodically. \textit{Chatbot personalization features} refers to whether firms use chat data to personalize users' conversations. \textit{Allows accounts for children 13-18} refers to whether the firm allows minors to use its chatbot.}
\label{tab:company-comparison}
\end{table*}

\section{Analysis}

We analyzed 28 documents across six chatbot developers, including privacy policies, linked sub-policies, and associated FAQs and guidance accessible from chat interfaces. Table 1 provides an overview of our key policy findings, while Table 2 provides an overview of the data sources used for training. 
In this section we review what we found in the policies: the sources of data used for model training, practices related to data retention, data sharing policies, human review of chat data, data de-identification, children's data, and personalization.

\textbf{Chat data for training.} We found that each of the six developers make use of user inputs to chatbots and corresponding outputs to train their AI systems. Until September 2025, Anthropic’s privacy policy stated that user inputs to Claude and corresponding outputs will only be used for model training if the user “explicitly opted in to the use of (their) Inputs and Outputs for training purposes” \citep{anthropic2025privacyupdate}. Anthropic announced in August 2025 that they were switching to opt-out by default.  OpenAI and Microsoft disclose that user data \textit{may} be used for model training while offering pathways to opt-out \citep{openai2024privacy, microsoft2025privacy}. Google and Meta, in contrast, state that users’ interactions with chatbots \textit{could} be used for model training and do not share clear routes to opt-out \citep{google2024privacy, meta2025genprivacy}. 
Amazon's privacy policy and the AWS privacy policy are silent on issues of AI training and chat data collection, though the Nova interface provides a notice that ``Your interactions and related information, including any content you submit like files or images, may be reviewed and retained'' to ``provide, develop, and improve our services, including AI models.'' Given the lack of detail provided in Amazon's written policies, we relied upon this statement to inform our analysis.


\textbf{Additional user-provided data sources.} 
Developers draw on other sources of user-provided data for training in addition to chat data. Amazon, Google, and OpenAI state that they may train on files that users upload with their prompts, while Microsoft explicitly excludes the contents of uploaded files. Microsoft and Google note that they may train on voice data, and OpenAI, Google, and Microsoft also state that they may train on user-uploaded images, though Microsoft notes that it attempts to de-identify images by removing ``metadata or other personal data and blurring images of faces'' \citep{microsoft2025privacy}. Microsoft and Meta notes that they may use a user’s data from their other products, in combination with input or output data, as part of model training as well. Microsoft’s privacy policy reads: ``Microsoft uses data from Bing, MSN, Copilot, and interactions with ads on Microsoft for AI training’’ \citep{microsoft2025privacy}. Meta's privacy policy states ``We also share information with Meta Companies to support innovation. For example, your videos can help train our products to recognize objects, like trees, or activities, like when a dog chases a ball’’ \citep{meta2024privacy}, while its generative AI privacy page notes ``we also use information shared on Meta products," including posts, photos, and photo captions, but not private messages \citep{meta2025genprivacy}. Despite the use of ambiguous language (e.g., "may") to describe their practices, unless explicitly disclaimed, these companies are training on chat conversations that broadly include detailed snapshots of their users' personal information, as personal content from photos, posts, and private chats with an AI assistant may be included in subsequent training runs.

\textbf{Sensitive Personal Data.} 
A significant privacy challenge chatbots pose is that not only may users disclose any type of personal information to a chatbot, including sensitive personal data, but also they may do so with additional context often absent in other forums. For example, a fact about one's health status is considered sensitive data, but disclosing it in the context of a conversation where one is seeking knowledge about or assistance with their health status can be far more revealing (e.g., in disclosing the cause of the ailment, such as the difference between knowing someone is HIV+ versus that they are HIV+ from intravenous drug use). The CCPA provides a list of data categorized as sensitive, which carry specific obligations for data processors.\footnote{The CCPA defines sensitive personal information as a subset of personal information that includes government identifiers (such as SSNs); account logins; specific financial account information; precise geolocation; contents of email and text messages; genetic data; biometric information processed to identify a consumer; information concerning a consumer’s health, sex life, or sexual orientation; or information about racial or ethnic origin, religious or philosophical beliefs, or union membership. 
}

In the policies we reviewed, only Microsoft acknowledged efforts to strip specific forms of personal data from user inputs to chatbots, stating that it removes ``information that may identify you, like names, phone numbers, device or account identifiers, sensitive personal data, physical addresses, and email addresses, before training AI models" \citep{msft_privacy_faq_copilot}. Anthropic and OpenAI take a different approach, acknowledging that while there may be personal data from the web within their training corpora, their rules-based approach to training their models includes privacy by design; Anthropic notes it trains its models to ``not disclose or repeat personal data which may have been incidentally captured in training data, even if prompted'' \citep{Anthropic_model_train, openai2025howchatgpt}. It is possible that other developers have similar practices of proactively removing identifiable and sensitive data types from their users' chat data, but they were not disclosed in the documents we reviewed, begging the question of where this practice should be disclosed to users if not in the privacy policy. While stripping such data from chats is helpful for mitigating certain privacy risks, it does not resolve privacy issues given possible contextual disclosures in a chat session and LLMs' ability to make inferences about users' personal information \citep{staab2024memorizationviolatingprivacyinference}.

\textbf{Default opt-in vs. opt-out for training on chat data.} After Anthropic's switch from opt-in to opt-out, all six of the developers now train on their users' chat data by default. 
A lack of affirmative consent from users for the use of their chat data in model training raises questions about how users may manage chats that contain sensitive data or topics, particularly if they do not or are unable to delete them. 
\newline
\indent The incorporation of user data for model training by default is a practice unique to consumer products. Based on our analysis, enterprise users' data is typically excluded from model training. For example, OpenAI states that ``Unless they explicitly opt-in, organizations are opted out of data-sharing by default’’ \citep{openai2024privacy, openai2025datacontrols}.  These practices create a two-tiered system where chat privacy is either unavailable or requires deliberate opt-outs for the vast consumer user base while paying customers have it by default.
\newline
\indent \textbf{Children’s data.} Developer practices varied regarding whether they collected children's data and how they used it. Unlike other developers, Anthropic neither collects any data from nor accepts any users under the age of eighteen, though they do not require users to age verify before creating accounts. Microsoft collects data from children under eighteen, but does not use it to build language models, stating that ``We do not train Copilot on data from…[a]uthenticated users under the age of 18’’ \citep{microsoft2025privacy}. Google, on the other hand, expanded its Gemini chatbot user base to include children under thirteen in May 2025 \citep{nyt2025geminikids} and will train its LLMs on data from children aged 13-18, but only if those children opt-in \citep{google2024privacy}. Amazon, Meta, and OpenAI do not explicitly note their approach to this issue, though all three developers allow users thirteen and older to create accounts, indicating that they do not treat older children’s data differently and therefore likely train on it by default. Training on children's data raises consent issues, as children cannot legally consent to it. Further, there is increasing evidence that children are using chatbots for parasocial, sexualized relationships that strain ethical (and potentially legal) boundaries \citep{Horwitz-meta}.
\newline
\indent \textbf{Data retention.} Every chatbot developer that we analyzed appears to retain some chat data indefinitely, primarily for trust and safety investigations or quality assessments.  Data deleted at specific points in time are often removed because they are a particular type, or because of a user action, such as marking a chat for quality review. For example, Google's default data retention term for Gemini is 18 months, though users can set it for three or 36 months instead \citep{google2025geminiprivacy}. However, their default retention period for chats reviewed by human reviewers (disconnected from one's Google account) is three years. Anthropic's data retention period is thirty days for users who have opted-out of using their chats for training purposes, but five years otherwise. Developer justifications for indefinite data retention are also often vague---for example, Meta states: “We keep information as long as we need it to provide our Products, comply with legal obligations or protect our or other’s interests” \citep{meta2024privacy}. Even in cases where data is not specifically retained, it is possible that the developer continues to make use of it in AI models that incorporated the data during earlier training \citep{cooper2025filescomputercopyrightmemorization}.
\newline
\indent \textbf{Human-review of chats.} Google and OpenAI's policies discuss using humans to review user chats for model training. Both companies state that users should not enter personal information into their chats that they would not want others to view, with Google specifically noting: ``Please don’t enter confidential information in your conversations or any data you would not want a reviewer to see or Google to use to improve our products, services, and machine-learning technologies" \citep{google2025geminiprivacy}. Such policies highlight the challenge with deidentifying chats, and raise questions about the steps developers take, if any, to attempt to deidentify chats beyond delinking them from a user's account information. For example, reporters interviewed contract workers reviewing chat transcripts on behalf of Meta who disclosed that they routinely had access to identifiable information from customer chats; the reporters were able to positively identify at least one individual from chat data shared with them \citep{bizinsider}.
\newline
\indent \textbf{Data de-identification.} The developers vary in the degree to which they explicitly commit to de-identify personal information before it is used for model training. Microsoft, OpenAI, and Anthropic suggest they may de-identify personal data used in model training. OpenAI notes “we may also aggregate or de-identify Personal Data so that it no longer identifies you and use this information for the purposes described above, such as to analyze the way our Services are being used, to improve and add features to them, and to conduct research” \citep{openai2024privacy}. Relatedly, Anthropic states: “We may process personal data in an aggregated or de-identified form to analyze the effectiveness of our Services, conduct research, study user behavior, and train our AI models as permitted under applicable laws” \citep{anthropic2025privacy}. 
Conversely, Amazon, Meta and Google do not state that they make any explicit exclusions of personal data from model training, and do not state whether any personal data used for training will be de-identified. Approaches to de-identification may instead be addressed in the context of specific model launches (e.g., as detailed in a technical report), though these practices for API models may not apply for black-box models that power chatbots.
\newline
\indent \textbf{Long-term personalization.} Google, Meta, Microsoft, and OpenAI  now offer personalized experiences with their chatbots that persist beyond a specific chat session. At the time we conducted this study, only Microsoft provided details linked from its privacy policy about how personalization functioned in Copilot, which was enabled by default and recalled ``key details you share, such as your name, interests, and goals" across conversations \citep{msft_privacy_faq_copilot}. Personalization is also on by default for OpenAI, which states in an FAQ that ChatGPT ``can recall details and preferences you’ve shared and use them to tailor its responses...ChatGPT can also use memories to inform search queries when ChatGPT searches the web using third-party search providers'' \citep{openai2025memory}.

\begin{table*}[t]
\centering
\small
\setlength{\tabcolsep}{8pt} 

\resizebox{\textwidth}{!}{
\begin{tabular}{|l|c|c|c|c|c|c|}
\hline
 \textbf{Data Source} & \textbf{Amazon} & \textbf{Anthropic} & \textbf{Google} & \textbf{Meta} & \textbf{Microsoft} & \textbf{OpenAI} \\
\hline
User chat inputs/outputs & \NotSpecified & \Yes& \Yes & \Yes & \Yes & \Yes \\
\hline
User uploaded documents  & \NotSpecified & \NotSpecified & \Yes & \Yes & \No & \NotSpecified \\
\hline
User uploaded images & \NotSpecified & \NotSpecified & \No & \Yes & \NotSpecified & \NotSpecified \\
\hline
Human annotated inputs/outputs & \NotSpecified & \NotSpecified & \Yes & \NotSpecified & \NotSpecified & \Yes \\
\hline
User feedback & \NotSpecified & \Yes & \Yes & \NotSpecified & \NotSpecified & \NotSpecified \\
\hline
Publicly available web data & \NotSpecified & \Yes & \Yes & \Yes & \Yes & \Yes \\
\hline
Licensed data & \NotSpecified & \Yes & \NotSpecified & \Yes & \NotSpecified & \Yes \\
\hline
Platform data & \NotSpecified & \NotSpecified & \Yes & \Yes & \Yes & \No \\
\hline
\end{tabular}
}
\caption{Sources of Training Data Across Developers. Each column is a chatbot developer and each row is a source of data. `Yes' signifies that the firm's privacy policies explicitly indicate it trains AI models on that source, `No' signifies it explicitly states it does not, and `Not Specified' indicates that either the developer's policies do not address that source or are ambiguous.
}
\label{tab:company-comparison}
\end{table*}

According to \citet{sampson2025personalization}, ``[i]mplementation of long-term memory can range from explicit user-guided storage, where individuals directly specify what should or should not be remembered, to system-driven approaches that dynamically recognize and store key information without explicit instruction as a user interacts with their AI assistant." Core privacy questions with respect to both personalization and long-term storage of memories or chats are whether such data will be excluded from model training, whether it will be shared across product boundaries (which may violate the principle of purpose limitation) or with third parties, and within the context of the CCPA, whether such data can be accessed, corrected, and deleted by consumers. When the data are used to generate inferences about specific individuals, as outputs, those too may be subject to the CCPA.

\textbf{Branch policies.}
We examined how companies disclose the use, collection and retention of chat data in their privacy policies, specifically focusing on the distinction between the main (parent) privacy policy and any subsidiary (branch) policies, which are typically linked within the main privacy document and pertain specifically to the company’s AI systems. We relied on the CCPA's requirement that privacy policies must be comprehensive in their scope, and that all essential information users need to understand how their chat data is collected and used would be found in the main privacy policy. However, as we conducted our review, we repeatedly encountered instances of pointers within main privacy policies to product-specific sub-policies or product FAQ documents that contained additional information about data collection and use practices not included in the main policies. We collected and coded these additional documents for each company, and then compared branch policies with the main policy for each company.

Most of the companies we analyzed have supplemented their main privacy policy with specialized branch policies for a variety of their AI products. These branch policies usually address specific considerations, jurisdictions, and features of the product, and in the case of AI models, they often focus on model-specific training and data handling practices, such as how user inputs contribute to model training, whether human reviewers are involved in the process of reviewing inputs, or whether a model is trained on deidentified inputs or on personal data from the internet. Google’s policy for Gemini, the Gemini Apps Privacy Notice, OpenAI’s policy for ChatGPT in its help center, and Anthropic’s Model Training Notice for Claude are all examples of branch policies that disclose AI-specific data practices that are not discussed in full (or in generalized terms) in the companies' main privacy policies. For example, for OpenAI we reviewed and coded six documents to capture their data collection and use practices: the main privacy policy along with its data processing addendum and four FAQs pages. That material facts related to data collection and use were disclosed not in a primary privacy policy document but across branching sub-policies raises questions about not only compliance, but also about the utility of privacy policies in the context of LLM-powered chatbots.

\textbf{Publicly Available and Licensed Data.} Five of the six developers acknowledge in their policies that they obtain publicly available data from the internet for model training purposes, and two (OpenAI and Meta) state that personal information may be included in crawled data. The one exception is Amazon, which makes no mention of AI training in its privacy policy. While each of the other five mention using some variation of publicly available information from the internet, Microsoft in particular notes that it uses data ``collected from industry-standard machine learning datasets and web crawls" \citep{msft_privacy_faq_copilot},  while Meta qualifies that it "get[s] datasets from publicly available sources, research institutions and professional and non-profit groups" \citep{meta2025genprivacy}. OpenAI provided the most detail, stating that it uses only ``information that is freely and openly accessible on the internet'' and that it excludes data behind paywalls or from the dark web \citep{openai2025howchatgpt}. Anthropic, OpenAI, and Meta all state that they license data from third parties for training purposes. Google's policies made no mention of this fact, though it was publicly reported that Google has licensed data from Reddit for this purpose \citep{schomer2024}. 

\section{Discussion}

Our findings highlight the ongoing tensions between principles and rights underlying privacy and data protection and the need for more data to continually train and update large language models. In this section we discuss the following themes from our analysis: the limitation of privacy policies to fully communicate AI-related data practices to the public; the increasingly blurred boundaries between chatbots and platforms' other products and how this enables increased data collection; the power of defaults to force consent for use of chat data for model training; the risks of indefinite data retention; and, who benefits from the development of LLM-powered chatbots and the incorporation of chat data. 


\textbf{Privacy policies: too much---but also not enough.} Despite years of research demonstrating that privacy policies fail to communicate privacy practices to consumers in a clear and comprehensive manner, they remain critical documentation for the public (including researchers) as a definitive source of what data companies collect and how they use it. 

As we analyzed our data, it quickly became clear that most privacy policies were anything but comprehensive when attempting to understand how AI developers were using consumers’ chat data. In nearly every case, there were data practices discussed in branch policies that were not disclosed in the main privacy policy. Some of these practices may be captured in other AI-specific documentation, such as model cards, but it is highly unlikely that consumers are familiar with model cards and would even know to consider this documentation as relevant. Privacy policies in general are already too long and complex for most consumers to navigate; forcing them to hunt across multiple documents to find answers about how this new technology processes their data makes information seeking even more time-consuming. 

Disclosing substantive data collection and use practices outside of the main privacy policy is disorganized at best and misleading at worst. 
The frequency of this practice highlights the challenges of making consistent disclosures in a rapidly evolving AI marketplace. 
For example, in some instances the sub-policy documentation we reviewed was directly linked from the chat interface written in clearer and simpler language than the privacy policy itself. User-friendly documentation that is linked directly from the point of interaction may be more salient to users seeking help or clarification \citep{10.1145/2808117.2808119}, but it does not solve the problem that one must traverse a network of branching sub-policies to fully explain a developer's relevant privacy practices. 

As researchers, it was challenging for us to work across these many documents to verify our research questions; for consumers it may be practically impossible. At minimum, a privacy policy should be redundant to any additional documentation provided to consumers. A company should reiterate its practices within FAQs or other user-centric documentation in order to highlight salient disclosures that may aid a user in making just-in-time choices about their data. Additionally, developers' privacy policies are changing rapidly---several were updated over the course of our study---and users may not receive adequate notice of changes, especially if the main privacy policy remains static but sub-policies are updated. Most companies email customers when their privacy policy is updated with substantial changes, but not when updating or issuing sub-policies.

In sum, privacy policies, already imperfect sources of truth, may not be fit for purpose for the data collection and use policies of LLMs. As we discuss in our recommendations, it may be time to expand requirements on the purpose and function of privacy policies if they are to support greater transparency for AI development \citep{bommasani20252024foundationmodeltransparency}.

\textbf{Blurring product boundaries.} It was challenging to clearly parse whether one's use of other products on a developer's platform (e.g., Google Docs) produces data used for model training elsewhere (e.g., Gemini). What is clear from our analysis is that the major platforms' legacy product boundaries are blurring, especially as companies begin to deploy AI agents that cross existing product boundaries to reference and organize users' emails, documents, and images \citep{reid2025aiinsearch}. 

These design decisions pose fundamental challenges to the data protection principles of data minimization and purpose limitation. In the pressure to create general-purpose AI systems, businesses are incentivized to decompartmentalize their user data as well as to collect data from as many sources as possible. Data minimization requires that developers not collect more data than actually needed to deliver a product or service, while purpose limitation requires that the use of data is tied to the reason given for its collection. Both are instantiated in the CCPA, meaning that developers will have to creatively work within its constraints while developing AI systems that fundamentally challenge these principles. One strategy developers adopt is to write policies and notices that generalize the collection and use of user data as to potentially encompass any product usage. Another, which appears imminent, is to integrate AI so broadly across a company's product offerings that every interaction point becomes a training data collection opportunity. 
 

\textbf{You \textit{will} improve the model for everyone: the power of defaults.}
The state of a system's defaults are one of the most critical design choices a company can make because defaults are sticky: most users do not change defaults, particularly if they impact a system-level choice (e.g., data flows) that is not visible in the UI \citep{nudge}. This is why default collection of chat data for model training by all developers has far-reaching implications for data privacy. 
Most users will not opt-out unless clearly prompted or the barrier to do so is low. OpenAI specifically frames the option in its interface in terms of social benefit: “Improve the model for everyone: Allow your content to be used to train our models, which makes ChatGPT better for you and everyone who uses it'' \citep{openai2025datacontrols}. The framing of this appeal in social terms rather than individual (e.g., `improve the model', or `improve the model for you') is designed to invoke users' guilt, also known as \textit{guiltshaming} \citep{10.1145/3173574.3174108} and attempts to persuade users by aligning the company’s interests with the public good \citep{10.1145/3359260}.

Opt-ins by default are at odds with the principle of consent. Users today are unlikely to understand what they are consenting to when allowing companies to train on their chats. The conversational nature of chatbots and large context windows encourages greater disclosure of personal information than search interfaces \citep{king2018privacy}, not even accounting for the additional data sources users may upload to a chat. Clouding this data exchange is the reality that it is practically impossible to isolate the impact of any particular piece of training data, making estimating the risk from inclusion challenging, though research suggests it is nonzero \citep{ nolte2025machinelearnersacknowledgelegal}. The breadth of personal data in training datasets can allow LLMs to make inferences about individuals who are not users of LLMs and to extrapolate based on novel text or image inputs \citep{staab2024memorizationviolatingprivacyinference, tömekçe2024privateattributeinferenceimages}
Such capabilities combined with web-scale training data may mean the death of the right to be forgotten \citep{zhang2024rightforgotteneralarge, tpp-rtbf}.

\textbf{Data leakage.} 
LLMs are known to memorize and regurgitate personal data, whether inadvertently or due to adversarial attacks \citep{lee2024talkinboutaigeneration}. 
Companies' safeguards against such attacks have been consistently subverted, putting users at risk \citep{nasr2023scalableextractiontrainingdata}.
All but one of the companies in our sample (Amazon) stated in their privacy policies that their training data could (and likely does) include personal information scraped from publicly available sources, meaning the personal information of non-users can also be leaked as it is in the training corpus. 

The policies we reviewed provide no details regarding how any developer stores and processes user inputs (chat histories and memory) and chatbot outputs (reasoning tokens and responses), and whether all of these sources are used for model training barring opt-outs. Prior to being processed for model training, chatbot inputs and outputs are likely held in temporary storage, but developers' policies do not clarify when or if this data is unlinked from user identities. Developers must provide more transparency on how long inputs and outputs are accessible and deletable by consumers, as they are subject to CCPA rights requests \citep{duffourc2024privacy, UCLA-NYU}. 


\textbf{Risks of indefinite and long-term chat data retention.} Amazon, Meta, and OpenAI retain some, if not all, of their users' chat data indefinitely. While some policies have nuanced qualifications about how long data might be stored (i.e., when it is used to address trust and safety issues), others, such as Meta, appear to have no limit. Chat data may prove to reveal more personal information than search queries, especially when aggregated across sessions. Given the depth of personal disclosure that can exist in chats, their retention raises several concerns: they are a valuable target for hackers; a breach of chat data could pose grave privacy risks to users; and over time the chats provide developers with a detailed dossier of an individual's life experiences, thoughts, feelings, and preferences, which may be used by developers for purposes such as profiling \citep{burgess, app13116355}. Indefinite chat retention policies are reminiscent of Google's past policy of eternally storing search queries linked to one's account, which Google eventually changed in 2020 to an 18 month default \citep{Hern}. 

Whether chat data can be fully anonymized is debatable; 
chat data with personally identifiable information removed may still reveal details that can identify individuals. The inclusion of children's chat data in training sets risks creating extensive digital trails for youth, potentially violating California's children's digital erasure law \citep{ca-bpc-22581}. Mass data collection and indefinite detention of what users may consider to be private conversations poses major risks for civil litigation as well as surveillance by government or law enforcement \citep{gabriel2024ethicsadvancedaiassistants}. 

\textbf{In whose legitimate interest are models being developed?} 
Each of the major platforms have, at best, a poor track record with respect to responsible use of consumers’ personal data for non-AI related purposes. While fully analyzing our findings within the framework of GDPR is out of scope for this paper, it is worth noting that of the companies that referenced GDPR in their policies, all made claims to legitimate interest as the basis for their data collection. At a high level, legitimate interest generally applies (with exceptions) when data processing is necessary to carry out a business activity \citep{legit_int}. However, the rights and freedoms of individuals cannot be seriously impacted by the business activity---simply creating a business case that relies on data collection that widely violates privacy is not a legitimate justification for the practice. This is why, for example, European data protection authorities have prohibited Clearview AI, a U.S. company, from doing business across the EU: because its business model relies upon collecting biometric facial data without individual consent, its legitimate interest is outweighed by EU citizens' rights and freedoms under law \citep{clearview}.
\newline
\indent Because LLM developers' business models are reliant on casting a broad net for data, including potentially collecting personal and identifiable non-customer data from the internet, their justifications for these practices have focused on claims about AI's transformative potential for public good. 
In addition to OpenAI's invocation of societal benefit, Microsoft makes appeals to diversity and inclusion, writing: ``By using real-world consumer data to help train our underlying generative AI models, we can improve Copilot and offer a more reliable and relevant experience. The more diversity in conversations our AI models are exposed to, the better they will understand and serve important regional languages, geographies, cultural references, and trending topics of interest to you and other users.'' These customer-centered appeals frame the developers' needs in terms of a win-win: \textit{help us improve our products and we both will benefit}. Unstated, of course, are the potential risks to users, which could help assess if companies' using chat data is truly in users' best interests.
 
\section{Recommendations}
Below we present five recommendations to address the concerns raised in our discussion. These recommendations do not preclude the need for a larger set of technical interventions (e.g., federated learning, differential privacy, synthetic data) and improvements in the transparency of training data. 
\newline
\indent \textbf{Adopt comprehensive federal privacy regulation.}
Without baseline consumer privacy regulations that put limits on the collection and use of personal data, the governance of personal data used to train LLMs in the U.S. will remain limited and fragmented. The CCPA provides a set of rights once personal data has been collected, but it excludes publicly available information. The conversational nature of chats may create difficulties for clean applications of CCPA, particularly when privacy concerns arise from aggregating chats and making inferences based upon them. Further, despite the inclusion of data protection principles like data minimization and purpose limitation in the CCPA, it may prove difficult for regulators to enforce such laws when developers create AI products that are built using as much data as possible. Federal privacy legislation should address the key data protection gaps in the U.S., including those we have raised related to LLM-powered chatbots \citep{king2024rethinking, solove2025}.

\textbf{Require opt-in for model training.} Policymakers should require developers to ask users to opt-in for model training rather than making it the default. With Anthropic's abandonment of opt-in as of September 2025, all six developers now train on user inputs by default. Children should never be opted-into data collection by default. In contrast, enterprise users are opted-out of model training by default, creating a two-tiered system where businesses have greater privacy protections over their data than the public. 

\textbf{Incorporate AI systems into privacy policies and release new transparency artifacts.} Our findings suggest that developers do not provide consumers with comprehensive information regarding how their personal data is used to train LLM-powered chatbots. 
Lengthier and more detailed disclosures, however, are not a solution. Developers must clarify how they collect, store, and process data for training models, both from user chats and the web, adopting standardized and user-friendly methods for communicating these practices. In addition to revamping privacy policies, companies should maintain robust, user-friendly AI-specific documentation, such as datasheets \citep{gebru2021datasheetsdatasets}, model cards \citep{Mitchell_2019}, and foundation model transparency reports \citep{bommasani2024foundationmodeltransparencyreports}. A privacy-focused version of Anthropic's Transparency Hub may be a potential model for developers \citep{anthropic2025transparency}. 


\textbf{Filter personal information from inputs by default.} Mitigation strategies for removing the low-hanging fruit of sensitive personal data from chats (e.g., social security numbers, health data) are the minimum intervention developers should implement irrespective of their data retention policies. Developers should proactively filter personal data from chat inputs by default, including any files uploaded by users. Chatbot user interfaces should also make it simple for users to edit, delete, filter, or otherwise strip information from their chats. 
In our review, only Microsoft's policies explicitly stated that it attempts to identify and remove some forms of personal data from chats. Other developers may curate data before it becomes part of the training corpus, but they do not document this in their privacy policies; developers should improve documentation and release tooling that makes such filtering broadly easier for data curators \citep{soldaini2024dolmaopencorpustrillion}.

\textbf{Promote innovation in privacy-preserving AI.} Short of a full stack privacy by design approach, developers can improve data protection through privacy features. OpenAI and Google offer users the ability to create ``Temporary Chats'' that are not used to train models and that are stored for shorter periods (30 days for OpenAI, 3 days for Google) \citep{openai2025datacontrols, google2025temp}. 

Other companies have adopted a variety of pro-privacy approaches. For example, Apple does not employ user data to train its AI systems \citep{apple_foundation_models_introducing_2024}; Proton has a no-logs policy on user chat inputs or chatbot outputs \citep{proton_lumo_privacy_2025}; and Tinfoil uses secure hardware enclaves to serve AI models in a verifiably private way \citep{tinfoil_enclaves_overview_2025}. The Allen Institute for AI has developed language models that allow data owners to decide when their data is active in the model \citep{shi2025flexolmoopenlanguagemodels}. Open-weight models offer major privacy benefits, as users can run models on their local machines without providing data to a third party \citep{kapoor2024societalimpactopenfoundation}. Frontier AI developers should adopt and built on these innovations rather than making user privacy an afterthought.

\section{Conclusion}
In the thirty-plus year span of the commercial internet, U.S. policymakers have been unable to adopt data privacy regulation that provides meaningful limits on or control over the personal data collected by companies. As chatbots are poised to fundamentally change how we disclose personal data not just for our information seeking needs but also to find answers to highly personal and intimate questions, limits must be put in place to prevent firms from exploiting this data. While state laws like the CCPA have some impact on data privacy, they are not sufficient for the sea change resulting from AI adoption.

As LLMs continue to develop at a breakneck pace, the privacy of our chat data is broadly at risk. A core question we must weigh is whether potential gains in AI capabilities from training on chat data are worth the loss of privacy and increased risk of domination of consumers. If these promises are to be realized, it must be based on a transparent and trusting relationship with the public. If people believe they are being exploited and manipulated for their data to benefit tech companies above all, then the public will be reluctant to accept the tradeoffs no matter the potential societal gain.


\bibliography{aaai25}
\end{document}